\documentclass{frontiersSCNS}

\usepackage{url,hyperref,lineno,microtype,subcaption}
\usepackage[onehalfspacing]{setspace}
\usepackage{url}

\usepackage{xcolor} 
\hypersetup{                              
    colorlinks = true,                    
    citecolor = {blue},
    linkcolor = {green},
    urlcolor  = {purple},
           }

\graphicspath{{./}{./figs/}}

\def\keyFont{\fontsize{8}{11}\helveticabold }
\def\firstAuthorLast{Martin {et~al.}}
\def\Authors{
Laura Martin\,$^{1}$, Bulcs\'u S\'andor\,$^{1,*}$ and 
Claudius Gros\,$^{1}$}
\def\Address{
$^{1}$Institute for Theoretical Physics, 
Goethe University Frankfurt, 
Frankfurt am Main, Germany}
\def\corrAuthor{Bulcs\'u S\'andor}
\def\corrEmail{sandor[.]itp.uni-frankfurt.de}

\begin{document}
\onecolumn
\firstpage{1}

\title[Closed-loop robots driven by transient synaptic plasticity]
{Closed-loop robots driven by short-term synaptic plasticity: 
Emergent explorative vs.\ limit-cycle locomotion}

\author[\firstAuthorLast ]{\Authors}
\address{}
\correspondance{}
\extraAuth{}

\maketitle

\begin{abstract}
We examine the hypothesis, that short-term synaptic
plasticity (STSP) may generate self-organized motor
patterns. We simulated sphere-shaped autonomous robots,
within the LPZRobots simulation package, containing
three weights moving along orthogonal internal rods.
The position of a weight is controlled by a single
neuron receiving excitatory input from the sensor,
measuring its actual position, and inhibitory inputs
from the other two neurons. The inhibitory connections
are transiently plastic, following physiologically
inspired STSP-rules.

We find that a wide palette of motion patterns are
generated through the interaction of STSP, robot, and 
environment (closed-loop configuration), including 
various forward meandering and circular motions, together 
with chaotic trajectories. The observed locomotion is 
robust with respect to additional interactions with obstacles.
In the chaotic phase the robot is seemingly engaged in
actively exploring its environment. We believe that our 
results constitute a concept of proof that transient synaptic 
plasticity, as described by STSP, may potentially be 
important for the generation of motor commands and 
for the emergence of complex locomotion patterns, adapting
seamlessly also to unexpected environmental feedback.

We observe spontaneous and collision induced mode switchings, 
finding in addition, that locomotion may follow transiently 
limit cycles which are otherwise unstable. Regular locomotion 
corresponds to stable limit cycles in the sensorimotor loop, 
which may be characterized in turn by arbitrary angles of 
propagation. This degeneracy is, in our analysis, one of the 
drivings for the chaotic wandering observed for selected
parameter settings, which is induced by the smooth diffusion 
of the angle of propagation.

\section{}
\tiny
\keyFont{ \section{Keywords:} closed-loop robots, short-term
synaptic plasticity, limit cycles, sensorimotor loop, 
self-organized locomotion, compliant robot} 
\end{abstract}

\section{Introduction}

It has been argued \citep{pfeifer2007self,aguilar2016review} 
that `robophysics', defined as the pursuit of the discovery 
of biologically inspired principles of self generated motion, 
may constitute a promising road for eventually achieving 
life-like locomotor abilities. Distinct principles 
such as 
predictive information \citep{ay2008predictive},
surprise minimization \citep{friston2011optimal},
chaos control \citep{steingrube2010self},
empowerment \citep{salge2014empowerment},
homeokinesis \citep{der2012playful},
cheap design \citep{montufar2015theory}, and
curiosity \citep{frank2015curiosity}
have been studied in this context.
Behavior, resulting from 
guided self organization \citep{prokopenko2009guided} or 
autonomous adaption \citep{chiel1997brain}, 
may be generated in addition through suitable 
synaptic \citep{der2015novel,der2016search} and 
intrinsic \citep{sandor2015sensorimotor}
plasticity rules. 

Here we point out, that complex dynamics may 
be generated through a transient plasticity
mechanism widely present in the brain. Short-term 
synaptic plasticity (STSP) 
\citep{fioravante2011short,regehr2012short}
is an activity induced transient modulation of the
synaptic efficiency, which may lead either to 
facilitating or to depressing behavior lasting
from a few hundred to a few thousand milliseconds.
STSP has been argued, besides others, to be relevant
or causal for working memory \citep{barak2014working}, 
for the facilitation of time sequences of alternating 
neural populations \citep{carrillo2015cell}, for motor 
control in general \citep{nadim2000role}, and for the
sculpting of rhythmic motor patterns \citep{jia2016short}
in particular. Plasticity mechanisms similar to
STSP have also been shown to allow for stable gaits 
\citep{toutounji2014behavior} in neural networks which 
are distinctively simpler than the ones used 
conventionally for bio-inspired controllers
\citep{schilling2013walknet}.

\begin{figure}[t]
\centering
\includegraphics[height=0.3\textwidth]{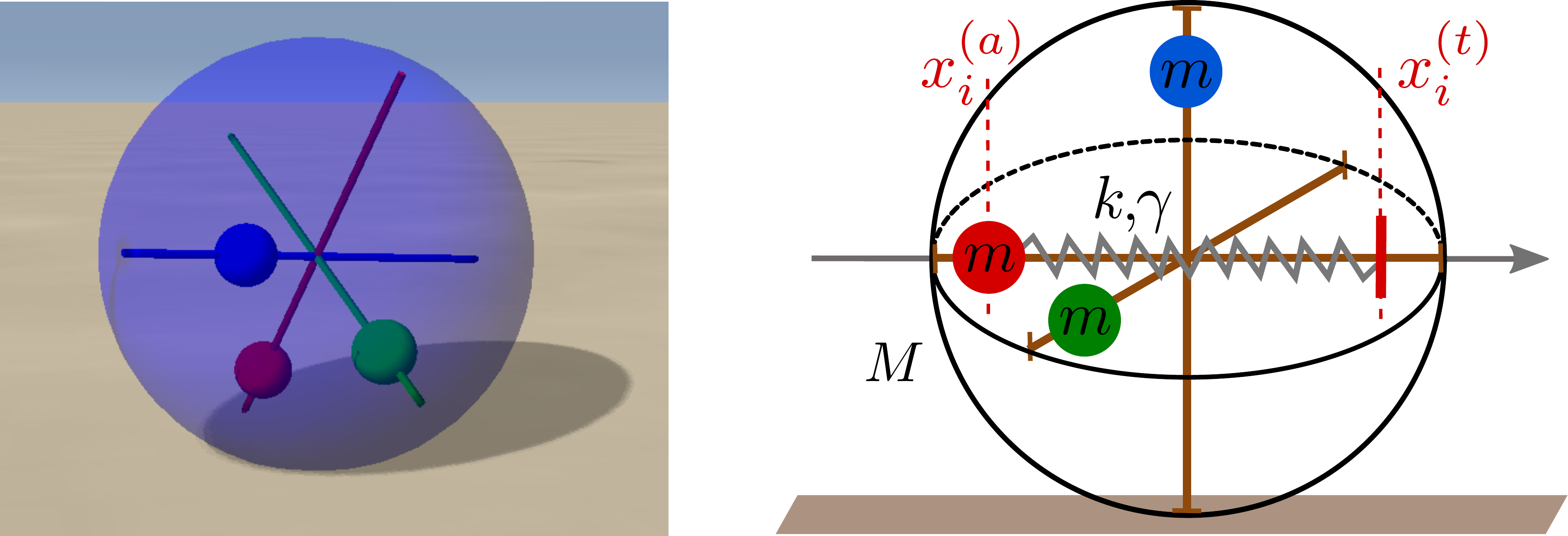}
\caption{\textit{Left:} A snapshot of the spherical robot 
from the LPZRobots simulation environment 
\citep{martius2013information}. The three weights (red, 
green and blue) can move along the respective rods without
interference
(\href{https://youtu.be/FSPXmmaWbDA}
      {click for movie}).
\textit{Right:} A sketch of the robot with 
the three perpendicular rods together with the 
three weights of mass $m$. The red vertical dashed 
lines show the actual position $x^{(a)}_i$ 
and a putative target position $x^{(t)}_i$ of the 
red weight along its rod. A damped spring with 
spring constant $k$ and damping $\gamma$ then
pulls the weight towards the target position, which
is given in turn by the output of a controlling
neuron (compare Fig.~\ref{fig:robot-control}).
}
\label{fig:three-axis-robot}
\end{figure}

In this study we use the LPZRobots physics simulation 
package \citep{der2012playful} for the investigation 
of the spherical three-axis robot illustrated in 
Fig.~\ref{fig:three-axis-robot}. This robot is driven 
exclusively by STSP, with locomotion coming to a 
stillstand both in the absence of synaptic plasticity and 
when the feedback from the environment is cut off, e.g.\ 
when the gravitational constant is set to zero. We find a 
surprisingly large palette of self-organized motion primitives, 
which includes a chaotic phase. The locomotion observed is 
flexible, in all modes, readjusting seamlessly to disturbances 
like the collision of the robot with obstacles. 

The capability of STSP to have a large impact on locomotion 
can be traced back in our analysis to the destabilizing effect 
short-term synaptic plasticity may have on attracting states
of the controlling network, inducing attractor-to-attractor 
transitions within timescales of the order of a few hundred 
milliseconds. We corroborate this findings by short-circuiting 
the sensori-motor loop, viz by taking out the environment. 
Transitions between distinct limit cycles within the full
sensori-motor loop are found in addition in the chaotic mode.

\section{Materials and Methods}

\subsection{Tsodyks-Markram model with full depletion}
The way neurotransmitters are released through the synaptic 
cleft may change transiently upon repeated presynaptic
activity \citep{tsodyks1997neural}, both for
excitatory \citep{wang2006heterogeneity} and for
inhibitory \citep{gupta2000organizing} synapses.
Physiologically this is, on the one side due to an
increase of the Ca-concentration $u \in [1,U_{max}]$ 
within the presynaptic bulge, facilitating the 
release of the respective neurotransmitter, and, 
on the other side, due to the decrease of 
the number $\varphi \in [0,1]$ of available 
vesicles of neurotransmitters. We use here with
\begin{equation} 
\begin{aligned}
\dot{u} &= \frac{U(y)-u}{T_u},&\qquad   U(y) &= 1 + (U_{max} -1) y \\
\dot{\varphi}& = \frac{\Phi(u,y) - \varphi}{T_{\varphi}},&\qquad  
\Phi(u,y)  &=1-\frac{uy}{U_{max}}
\end{aligned}
\label{eq:fullDepletion}
\end{equation}
a modified version of the original Tsodyks-Markram model 
\citep{tsodyks1997neural,hennig2015theoretical}, in which 
the the Ca-concentration
$u$ and the number of vesicles $\varphi$ of a given
synapse relax to target values $U=U(y)$ and $\Phi=\Phi(u,y)$,
determined in turn by the level $y\in[0,1]$ of the
presynaptic activity. A prolonged maximal presynaptic 
activity $y\equiv1$ would lead with $\varphi\to0$ to a 
full depletion of the reservoir of vesicles.

The dynamics of the full depletion model 
(\ref{eq:fullDepletion}) is determined by the
relaxation time constants $T_u$ and $T_\varphi$,
and by the maximal level $U_{max}$ of the
Ca concentration. For $U_{max}=1$ a monotone 
depression is present, whereas $U_{max}>1$ 
initially generates facilitation by a fast 
calcium influx, being annulled later on by 
the depletion of neurotransmitters. Overall, 
the synaptic efficiency is proportional
to $u\varphi$, viz to the number of vesicles
and to the release probability (which in turn
is assumed to be proportional to $u$). We use 
$T_u = 300\,\text{ms}$ and $T_{\varphi}=600\,\text{ms}$, 
together with either $U_{max}=1$ or $U_{max}=4$. These values
are within the typical range of what is physiologically 
observed \citep{wang2006heterogeneity,gupta2000organizing}.

\begin{figure}[t]
\centering
\includegraphics[width=0.8\textwidth]{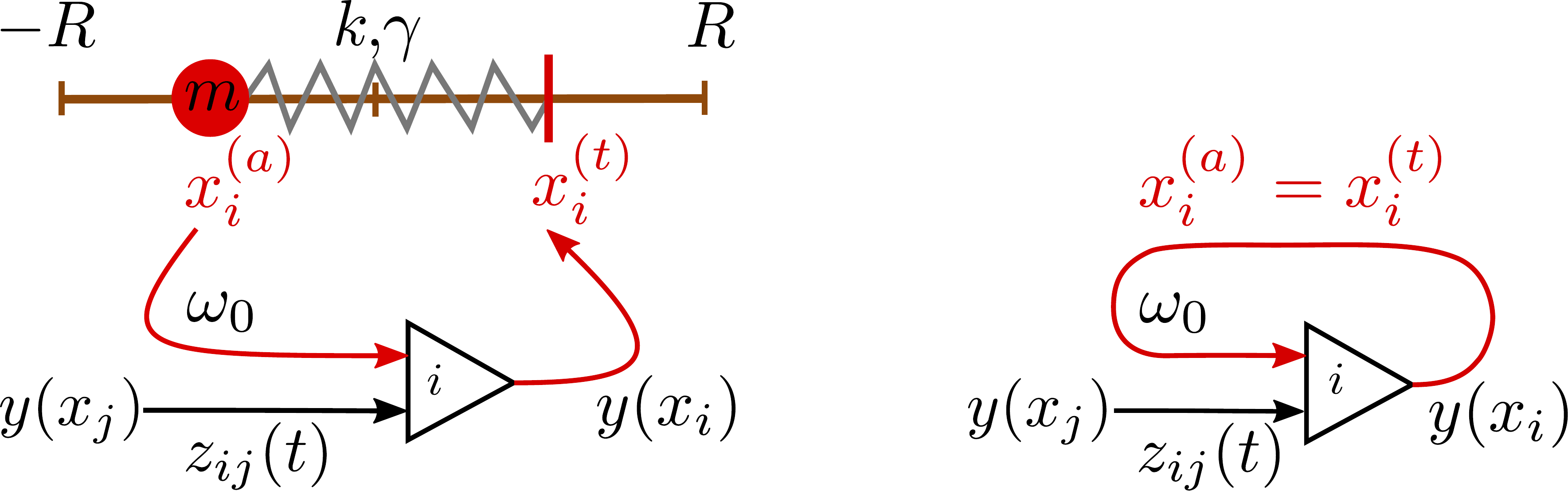}
\caption{\textit{Left}: Sketch of the sensorimotor
loop of the three-axis spherical robot illustrated
in Fig.~\ref{fig:three-axis-robot}. The three weights
$i=1,2,3$ with masses $m$ are each controlled by a 
single neuron. The excitatory input $w_0 (x_i^{(a)}+pR)/(2pR)$ 
of neuron $i$ is proportional to the proprio-sensory 
measurement of the actual position $x^{(a)}_i \in [-R,R]$
of the $i$-th mass ($p\in[0,1]$). The neuron also receives inhibitory 
inputs $-z_0\varphi_ju_jy(x_j)$ from the other two neurons 
($j\ne i$). The output $y(x_i)$ of the $i$-th neuron 
determines via $x_i^{(t)}=pR[2y(x_i)-1]$ the target 
position of the $i$-th mass.
\textit{Right:} A network of (three) neurons having
the identical topology as the one of the three-axis
spherical robot, but with the feedback of the
environment short-cut by identifying the actual 
position $x_i^{(a)}$ with the target position $x_i^{(t)}$.
}
\label{fig:robot-control}
\end{figure}

\subsection{The robot}
The movement of robot illustrated in 
Fig.~\ref{fig:three-axis-robot} is induced
by the relative gravitational pull of the
three weights, together with the rolling friction
and angular momentum conservation. The individual
neurons $i=1,2,3$ are modeled as rate-encoding
leaky integrators,
\begin{equation}
\dot{x}_i  = -\Gamma x_i + 
\frac{w_0}{2pR}\left(x_i^{(a)}+pR\right) 
- z_0 \sum_{j\neq i} u_j \varphi_j y(x_j),
\qquad\quad
y(x_j)=\frac{1}{1+\exp(-a x_j)},
\label{eq:dot_x}
\end{equation}
where $x_i$ and $y(x_i)$ are the respective membrane 
potentials and firing rates. $\Gamma$ is the relaxation
rate, $R$ the diameter of the robot, $p\in[0,1]$ a
rescaling factor, $x_i^{(a)}\in[-R,R]$ the sensory reading 
of the actual position of the weight on the rod, $w_0>0$ 
the weight of excitatory input and $z_0>0$ the magnitude of 
the inter-neural inhibitory connections. We note that
the variables of the STSP, $u_j$ and $\varphi_j$, as
described by (\ref{eq:fullDepletion}), depend only on 
the presynaptic activity and can hence be attributed
altogether to the presynaptic neuron. For the slope 
of the sigmoidal $a=0.4$ has been selected. The weight
of the excitatory input $w_0$ is not modulated here
by short-term synaptic plasticity, corresponding to
a direct sensory reading.

\begin{figure}[t]
\centering
\includegraphics[width=0.8\textwidth]{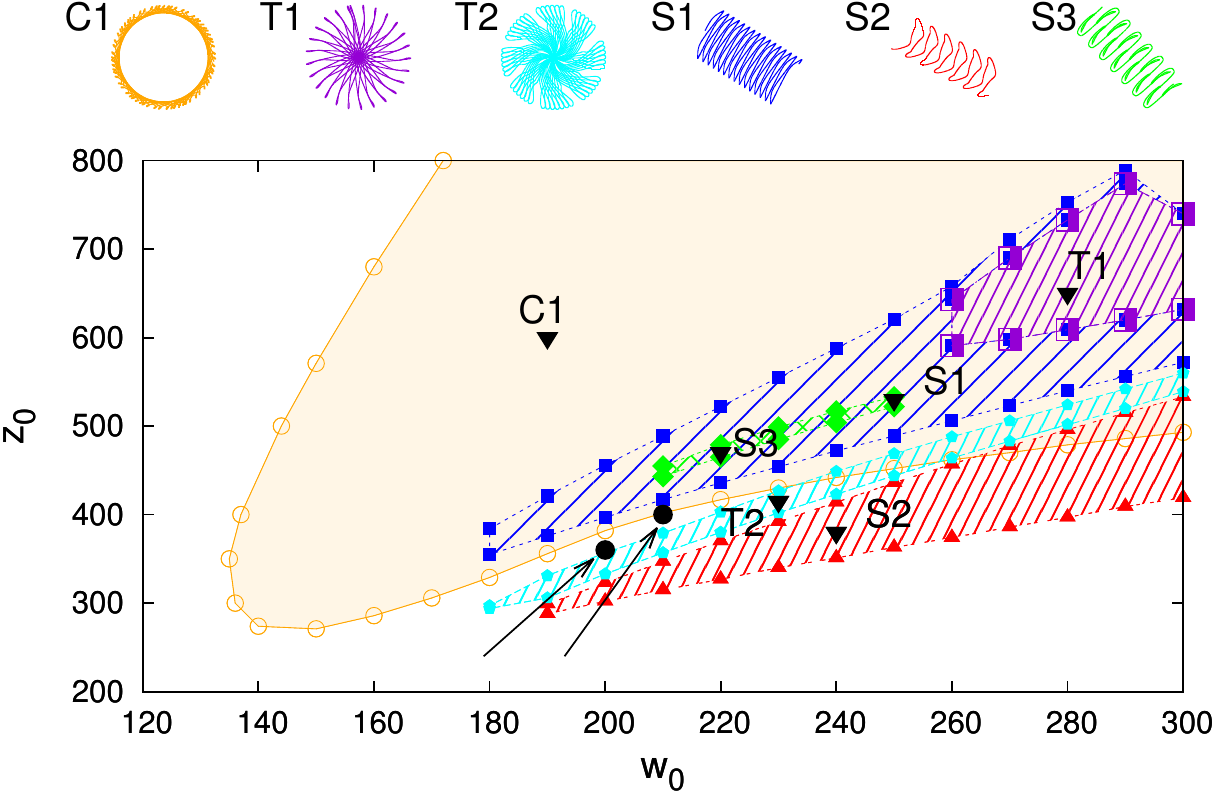}
\caption{Phase diagram for $U_{max} = 1$ in the parameter 
plane of excitatory ($w_0$) and inhibitory ($z_0$) synaptic 
weights. On the top the different types of identified regular 
motion patterns are illustrated, tagged respectively with black 
triangles in the respective regions of stability (shaded areas). 
Close-up trajectories are given in Fig.~\ref{fig:U1_modes}; 
for a comparison see also Suppl.~Video~1.
Examples of two parameter settings, (200,360) and (210,400), 
for which chaotic behavior is observed are indicated by black 
filled circles (at the tip of the respective arrows;
\href{https://youtu.be/t1hO16paIcs}
      {click for movie}).
}
\label{fig:U1_PD}
\end{figure} 

We selected with $p=1/2$ a reduced range for the 
target position $x_i^{(t)}$,
\begin{equation}  
\begin{aligned}
x_i^{(t)} = pR\left[2y(x_i) - 1\right],
\qquad\qquad
x_i^{(t)}\in[-pR,pR].
\end{aligned}
\label{eq:x_t}
\end{equation}
This choice allows to avoid dynamic overshooting of the 
weight when accelerated from its actual to the target 
position. The force accelerating the weight is calculated 
by the LPZRobots package by simulating a damped spring:
\begin{equation}
\begin{aligned}
m\ddot{x}^{(a)}_i  & = -k(x_i^{(a)}-x_i^{(t)})-\gamma\frac{d(x_i^{(a)}-x_i^{(t)})}{dt}+F_i,
\qquad\quad x_i^{(a)}\to x_i^{(t)},
\end{aligned}
\label{eq:damped_oscillator}
\end{equation}
where $k$ is the spring constant and $\gamma$ the 
damping. Centrifugal and other induced forces, $F_i$, 
act additionally in (\ref{eq:damped_oscillator}) on the individual weights. 
The complete setup of the three-neuron network is illustrated
in Fig.~\ref{fig:robot-control}. 

\subsection{Simulation parameters\label{sec:sim_par}}

The LPZRobots simulation environment \citep{der2012playful} 
is an interactive simulator based on the ODE (Open Dynamic 
Engine) \citep{smith2005open}. LPZRobots contains rigid body 
dynamics in terms of a library of basic primitive objects, 
such as spheres and cuboids, as well as a variety of joints, 
sensors and surface materials.

We used $roughness=0.8$, $slip=0.01$, $hardness=40$ and
$elasiticity=0.5$ for the collision and friction properties
together with $friction=0.3$ (the rolling friction coefficient),
$gravity=-9.81$ (the gravitational constant) and $noise=0$ 
(for the global noise level). All parameters are in SI units. 
For the stepsize of the physical simulation $simstepsize=0.001$
was used (corresponding to a millisecond).
With $controlinterval=1$ one ensures that the controller,
viz Eq.~(\ref{eq:dot_x}), is updated as often as the
physics of the environment.

The robot itself has a diameter of $2R=0.5$, a mass off $M=1$ 
and a $motorpowerfactor=120$. The parameters 
for the damped oscillator (\ref{eq:damped_oscillator})
are $m=1$, $k=m*motorpowerfactor$ and $\gamma=2\sqrt{k*m}$ 
(critical damping). The relaxation rate
for the membrane potential entering Eq.~(\ref{eq:dot_x})
has been set to $\Gamma=20$, retaining the bare excitatory
and inhibitory weights, $w_0$ and $z_0$, as free simulation 
parameters.

\begin{figure}[t]
\centering
\includegraphics[width=0.7\textwidth]{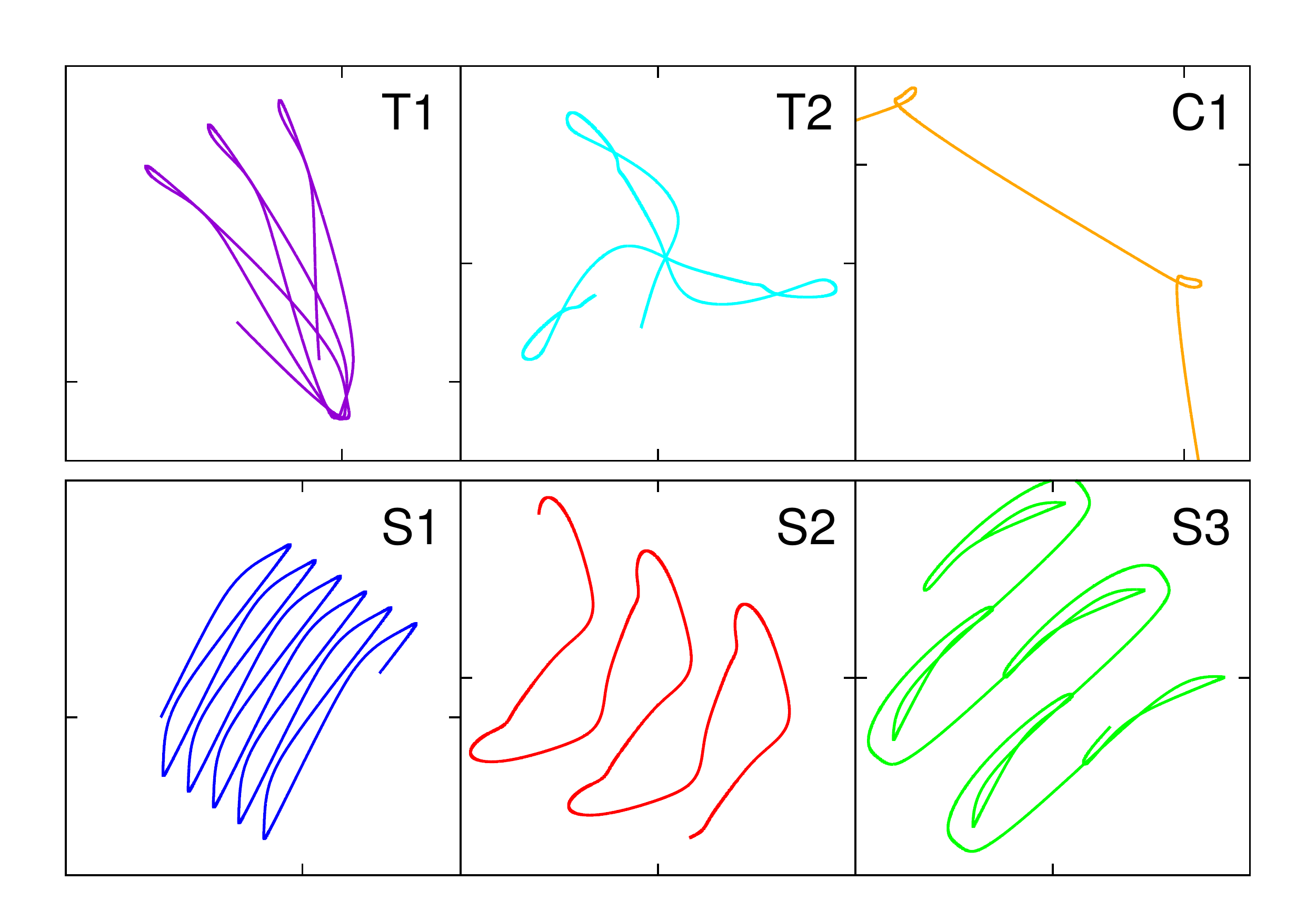}
\caption{A close-up of the trajectories in the plane of 
locomotion, for the parameters $(w_0,z_0)$ tagged as
black triangles in the phase diagram presented in
Fig.~\ref{fig:U1_PD}.
T1: $(280,650$), T2: $(230,415$), C1: $(190,600)$, 
S1: $(250,530)$, S2: $(240,380$), S3: $(220,470)$.}
\label{fig:U1_modes}
\end{figure} 

\section{Results}

\subsection{Emergent limit-cycle locomotion}

In Fig.~\ref{fig:U1_PD} we present the stability regions
for the various regular movement patterns found, with 
respective close-ups given in Fig.~\ref{fig:U1_modes}.
The results are for $U_{max}=1$ (depressing short-term 
synaptic plasticity without Ca dynamics) and for the 
parameters specified in 
Sect.~\ref{sec:sim_par}. They are obtained by adiabatically
continuing stable states along a grid until stability is lost. 
Without STSP only a globally attracting fixpoint corresponding 
to a motionless robot is present. We note that regular motion 
arises for a wide range of bare excitatory ($w_0$) and inhibitory 
($z_0$) synaptic weights. $z_0$ needs however to be larger 
than $w_0$.

All motion patterns observed are self-organized.
There is no objective function \citep{gros2014generating}, 
such as a maximal velocity, to be optimized. This implies
that the quantitative features of the individual motion patterns 
change smoothly within their respective stability regions, and 
that one can identify the observed regular movement patters as 
stable limit cycles in the sensorimotor loop 
\citep{sandor2015sensorimotor}. Fast switching between 
motion primitives would be possible by a putative overarching
controller, since more than one limit cycle may 
be stable for given synaptic weights $w_0$ and $z_0$.
Interactions between robots or with external obstacles 
might also lead to the automatic selection of another 
coexisting mode (see for instance Suppl.~Video~1).

\begin{figure}[t]
\centering
\includegraphics[width=0.75\textwidth,angle=0]{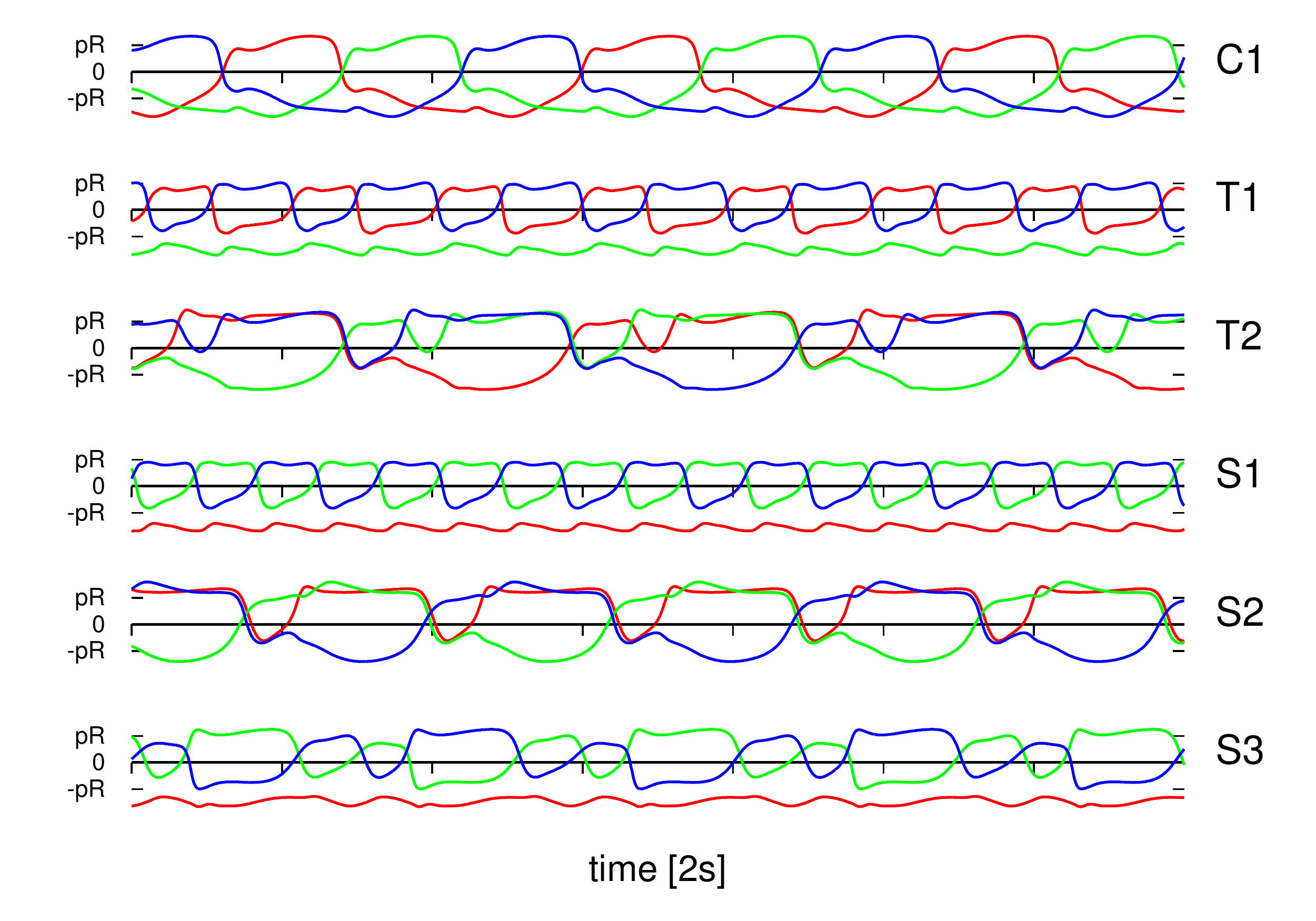}
\caption{The positions $x_i^{(a)}$ of the three weights as a 
function of time, compare Fig.~\ref{fig:three-axis-robot}, 
along the corresponding rods. The modes and parameters are
identical to the ones presented in Fig.~\ref{fig:U1_modes}.
Time is measured in units of $2\,\text{sec}$.}
\label{fig:modes_time_series}
\end{figure}

It is evident that the body plan of the robot
examined here tends to
produce meandering motion pattern. T1 and T2 are sun- 
and star-like movements with small (T1) and large (T2)
processing angles (compare Fig.~\ref{fig:U1_modes};
`T' stands for torus in phase space). There is, in 
addition, a (nearly pure) circular motion, C1, and 
three types of forward snake-like meandering 
motion patters, S1, S2 and S3. From these S3 partly 
overlaps with itself. These modes are characterized
by distinct motion patterns of the three weights,
as shown in Fig.~\ref{fig:modes_time_series}, as
measured by their positions along their respective
rods. The differences between the distinct modes
are in part qualitative, in terms of the time
sequences in which the three neurons are subsequently 
active, and in part only quantitative. The difference 
between T1 and S1 is, in this respect, that the up-times
of the two active neurons are symmetric for S1,
but not for T1. A spontaneous symmetry breaking can 
be furthermore observed in case of T1, S1, S2, S3, 
for which two weights always have alternating dynamics, 
the third one showing a qualitatively different behaviour.
In contrast to that, the time-series of the C1 and T2 
modes reveals the symmetrical but phase shifted 
oscillation of the three weights.
Note that the positions of the weights
may overshoot the interval $[-pR,pR]$ for the target 
positions $x_i^{(t)}$, both due to inertia and due to 
the additional gravitational pull. Motion patterns
similar to the ones shown in Fig.~\ref{fig:U1_modes}
have been observed in a self-organized two-wheeled
robot in the frozen mode \citep{der2013behavior}.

\begin{figure}[t]
\centering
\includegraphics[height=0.35\textwidth]{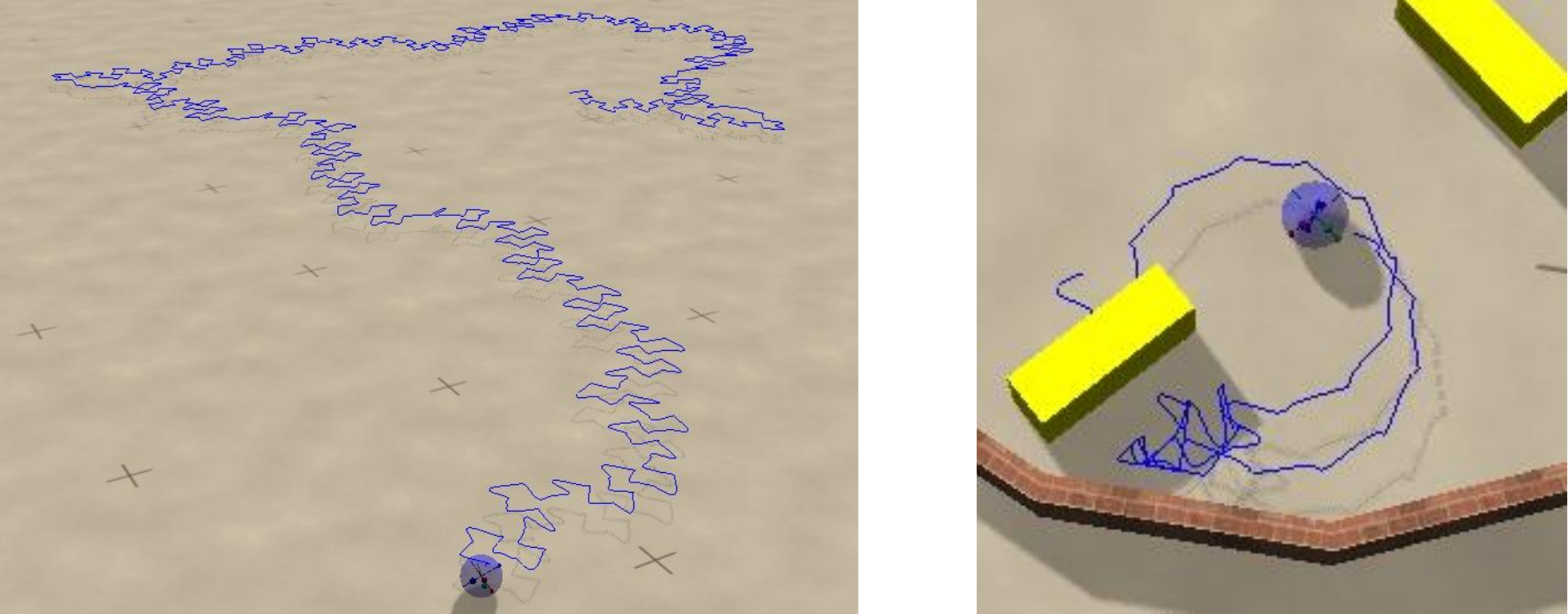}
\caption{Screenshots of the sphere robot in a chaotic mode;
$U_{max}=1$ and $(w_0,z_0)=(210,400)$. The blue lines retrace
the past trajectory. The short-time motion of the robot
is close to the one of the S2 mode, which is here an unstable
attractor (compare Fig.~\ref{fig:U1_modes}).
{\it Left}: In open space.
{\it Right}: In a closed environment allowing for the interaction 
             with movable objects (yellow blocks). The circular
             sections correspond to unstable C1 limit cycles
(\href{https://youtu.be/3eS0YQ5CawE}
      {click for movie}).
             }
\label{fig:U1_chaos_screenshot}
\end{figure}

\subsection{Chaotic modes allowing for explorative behavior}\label{SCM}

The dynamics of the robot takes place in a phase space combining
the internal variables, of both body and controller, 
with the ones of the environment. The stability regions of the
individual limit cycles presented in Fig.~\ref{fig:U1_PD} will
therefore be bounded, generically, by a suitable bifurcation,
such as a supercritical Hopf bifurcation or a fold bifurcation
of limit cycles \citep{gros2015complex,sandor2015sensorimotor}.
Alternatively, a transition to chaos may occur. It is on the
other side also possible that chaotic attractors emerge from 
previously unstable manifolds and that the stability region of 
chaotic and stable manifolds overlap.

Close to a chaotic phase long transients may occur, which 
makes it difficult to study systematically the exact extend 
of the chaotic region. In Fig.~\ref{fig:U1_PD} we have 
indicated however a few representative combinations of 
parameters, for which stable chaos is observed both in the 
limit of long simulations times and for a wide range of 
stepsizes of the ODE simulator. No regular motion patterns 
can be observed in the screenshots presented in 
Fig.~\ref{fig:U1_chaos_screenshot}. We have also evaluated 
the long-time behavior of the square of the covered 
real-space distance,
\begin{equation}
d^2(\tau) = \langle 
\left(\mathbf{x}(t+\tau)-\mathbf{x}(t)\right)^2\rangle_t~.
\end{equation} 
We found diffusive transport $d\sim\sqrt{\tau}$ for
the chaotic mode and ballistic transport $d\sim\tau$
for the forward meandering modes S1, S2 and S3. Both
as expected.

It has been observed, that chaotic locomotion of
an embodied system may be considered as a basic 
explorative behavior, both of the environment and 
of the own motor pattern 
\citep{shim2012chaotic,steingrube2010self}.
As a test of this hypothesis we have set our
three-rod robot into a restricted playground containing
movable objects in the form of blocks, which can
be pushed, to a certain extend, over the ground. A 
screenshot is presented in Fig.~\ref{fig:U1_chaos_screenshot}.
One can observe, that the robot stays for a while close 
to the object, bumping around, and retracting in part 
a trajectory having a shape similar to the one generated 
by a C1 limit cycle. This is possible, as the set of 
parameters $(w_0,z_0)=(210,400)$ considered is located 
close to but outside the C1-stability region. The C1 limit 
cycle is hence only weakly unstable in the chaotic phase.
The active exploration of the environment, occurring here
when bumping into obstacles, gives the robot hence access 
to otherwise unstable locomotion options. The overall
behavior may be interpreted alternatively in terms of
non-representational sensorimotor knowledge
\citep{buhrmann2014non}.
For a longer simulation see the supplementary videos.

In the movie presented in the supplementary
material one can observe, furthermore, that the
robot is pushing the blocks around in a seemingly
`playful' manner (see Suppl.~Video~3). 
A remarkable behavior, in our view,
considering that the sphere robot disposes of a mere
total of three controlling neurons. We note, that this
complex behavior results from the interplay of the 
autonomous dynamics, as resulting from the inter-neural 
short-term synaptic plasticity, with environmental 
feedback.

\subsection{Embodiment shaping the intrinsic dynamics}

\begin{figure}[t]
\centering
\includegraphics[width=0.70\textwidth]{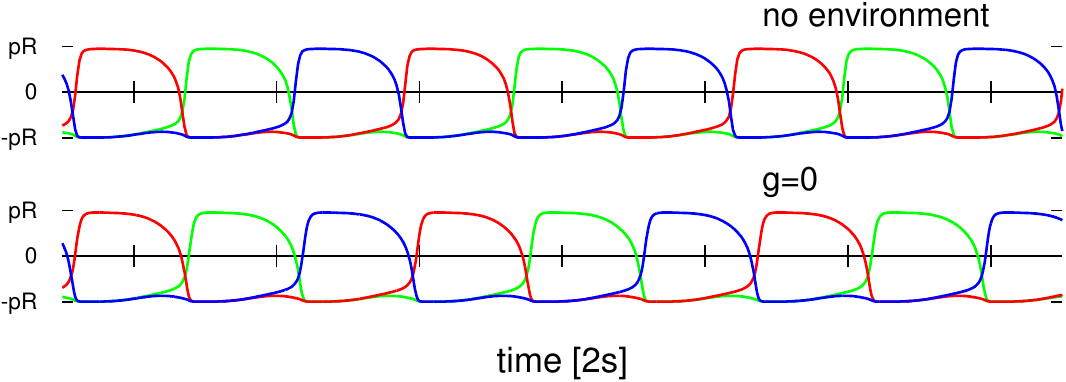}
\caption{Time series of the target positions $x_i^{(t)}$ 
for $U_{max}=1$ and $(w_0,z_0)=(190,600$),
which correspond to the C1 mode shown in Fig.~\ref{fig:U1_modes}
and \ref{fig:modes_time_series}.
{\it Top}: For a numerical simulation of the isolated 
network obtained when setting $x_i^{(a)}=x_i^{(t)}$ in
Eq.~(\ref{eq:dot_x}).
{\it Bottom}: For the 3-rod robot suspended in air (with 
the gravity constant $g$ set to zero). Note that both
time-series are very similar but not identical.
}
\label{fig:time_series_comparison}
\end{figure}

One can consider the controlling 3-neuron network 
in isolation by identifying the sensory reading 
$x_i^{(a)}$ for the actual position of the weight 
along the rod with the respective target position
$x_i^{(t)}$, viz by setting $x_i^{(a)}=x_i^{(t)}$ in
Eq.~(\ref{eq:dot_x}). The resulting network contains
a self-excitatory coupling $w_0$ together with all-to-all 
inhibition with a bare synaptic strength $z_0$. The 
short-term synaptic plasticity then induces an 
autonomous activity, as illustrated in 
Fig.~\ref{fig:time_series_comparison}, which
is topologically equivalent to the C1 mode. This
equivalence becomes even more pronounced when
suspending the robot in air, which can be achieved
in turn by simply removing gravity from the physics
simulation (bottom time-series in 
Fig.~\ref{fig:time_series_comparison}).
One can hence consider the C1 mode as the
driver for the observed physical motion.

\begin{figure}[t]
\centering
\includegraphics[width=0.80\textwidth]{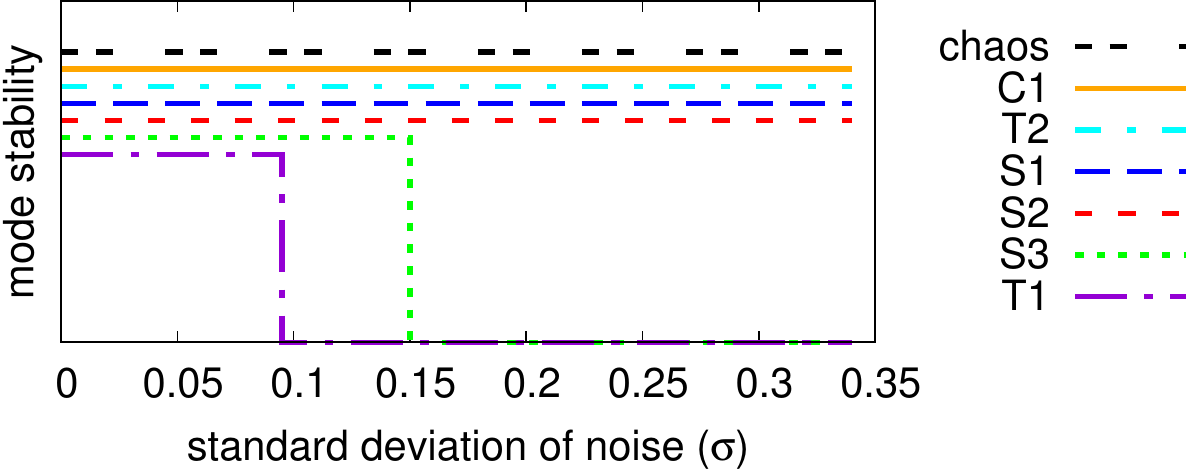}
\caption{Stability analysis of the modes found for $U_{max}=1$, 
compare Fig.~\ref{fig:U1_PD}, against a noise term $\Delta x$ in the 
sensory reading, defined by $x_i^{(a)}\to x_i^{(a)}(1+\Delta x)$.
Finite and zero values along the $y$-axis indicate stability and 
instability (the displacements along the $y$-axis are only for avoiding
overcrowding). The noise $\Delta x$ is normal-distributed with standard 
deviation $\sigma$. Once T1 and S3 become unstable, when adiabatically 
increasing the noise level, their respective basins of attraction merge 
with the attracting regions of C1 and S1.}
\label{fig:stability_vs_noise}
\end{figure}

The isolated 3-neuron network  has, however, only
a single stable limit cycle. Numerically integrating 
the isolated network for parameters settings $(w_0,z_0)$
corresponding to the six modes of 
Fig.~\ref{fig:modes_time_series}, as well as for 
chaotic states, we find always an identical sequential
activation of the three neurons illustrated in
Fig.~\ref{fig:time_series_comparison}, with only slight
changes in the overall shape. It is hence clear, that
the other modes T1, T2, S1, S2 and S3, as well as the
chaotic behavior, do result from the closed-loop feedback
of the environment. The interaction of the environment
with the intrinsic dynamics then results in the emergence 
of alternative types of locomotion.

\subsection{Stability with respect to noise}

We present in Fig.~\ref{fig:stability_vs_noise} 
an analysis of the stability of the various modes found,
with respect to noise in the sensory readings, where the
level of the noise is given by the relative standard deviation
$\sigma$ of the sensory readings $x_i^{(a)}$. Comparing with
the phase diagram, as presented in Fig.~\ref{fig:U1_PD},
one notices that first modes to disappear, T1 and S3,
are the ones with small stability regions in the phase
diagram. Ramping up the noise level the T1 and S3 
modes turn respectively, above their corresponding
critical noise levels, into C1 and S1 modes. The other modes, 
including the chaotic phase, are in contrast very stable 
with respect to noise.

\subsection{Autonomous mode switching}

\begin{figure}[t]
\centering
\includegraphics[width=0.80\textwidth]{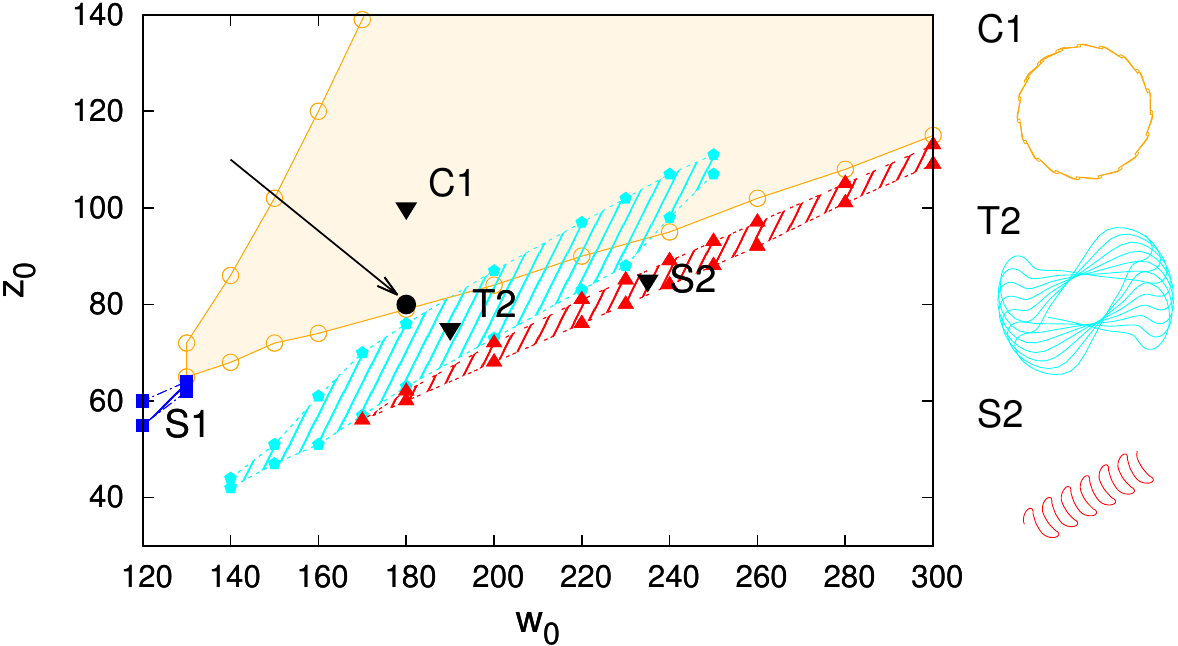}
\caption{The phase diagram obtained when using $U_{max}=4$ 
in the STSP rules (\ref{eq:fullDepletion}), with the naming 
of the modes corresponding to the ones used for the $U_{max}=1$ 
phase diagram presented in Fig.~\ref{fig:U1_PD}. At 
$(w_0,z_0)=(180,80)$ there is chaotic state, as indicated by 
the arrow, coexisting with the C1 mode. The extent of the
chaotic phase has not been examined in depth. On the right
the traces are shown for the three dominant modes C1, T2 and S2.
}
\label{fig:U4_PD}
\end{figure}

We present in Fig.~\ref{fig:U4_PD} the phase
diagram obtained when using $U_{max}=4$ for
the maximal Ca-level entering Eq.~(\ref{eq:fullDepletion}).
Within the range of $(w_0,z_0)$ scanned we find four
out of the six modes observed for $U_{max}=1$
(compare Fig.~\ref{fig:U1_PD}). The range of
inhibitory weights $z_0$ for which stable locomotion
is found is rescaled down, in addition, with respect
to the $U_{max}=1$ case. Interestingly we found
a chaotic state at (180,80) which lies just inside
the stability region of the C1 mode.

\begin{figure}[t]
\centering
\includegraphics[height=0.45\textwidth]{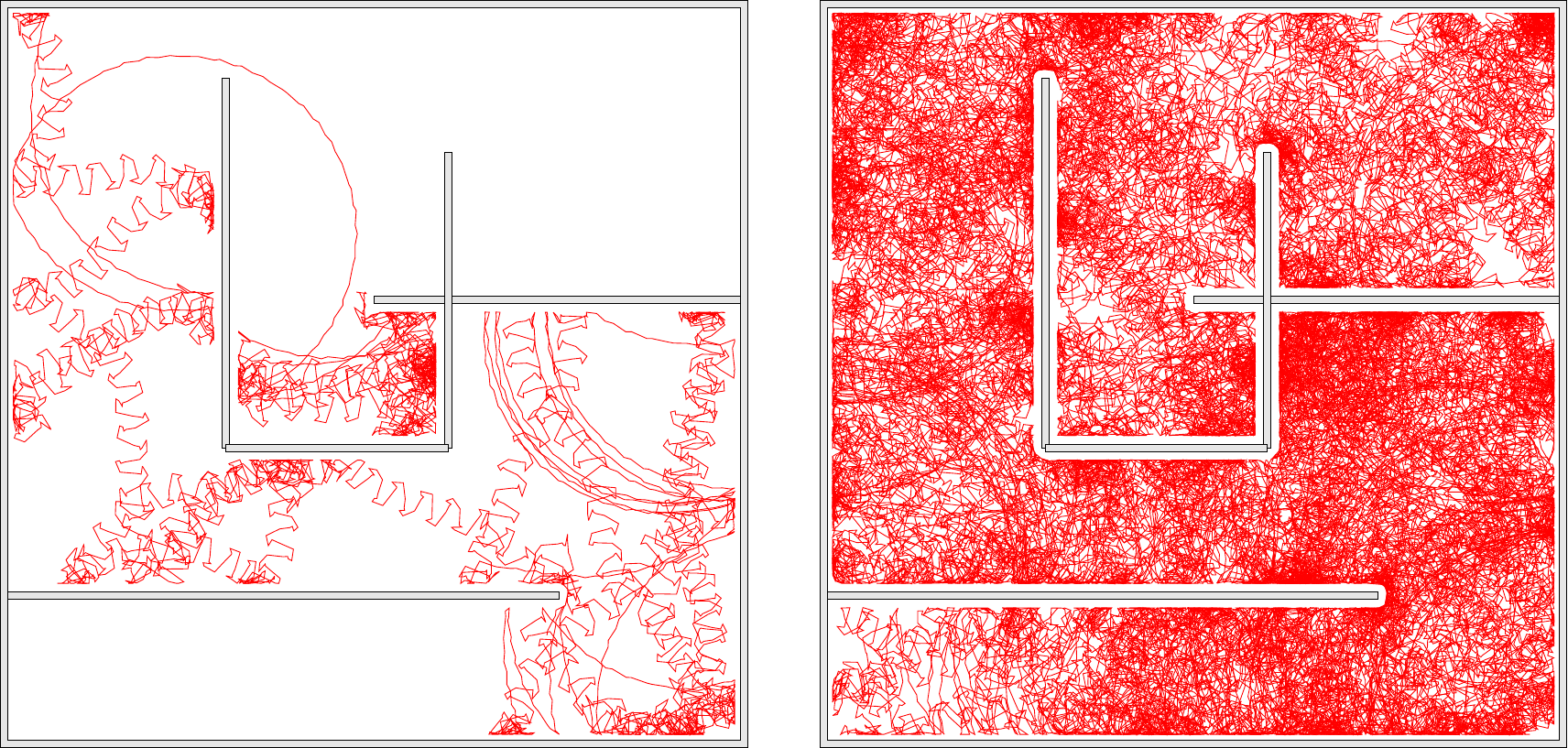}
\caption{The trace of the robot in a maze for a simulation
time of 83 (left) and 1000 (right) minutes respectively.
The robot may remain stuck occasionally in corners, but
not forever. The parameters are $U_{max}=4$ and 
$(w_0,z_0)=(180,80)$, corresponding to the chaotic mode 
indicated by the arrow in Fig.~\ref{fig:U4_PD}. Bumping 
against the wall the robot sometimes turns up in the C1 
mode, which is a coexisting stable limit cycle. The 
radius of the C1 mode is however, in this case, so large, 
that it does not fit as a whole into the maze. Also note 
that the chaotic mode is locally akin to the here unstable
S2 mode, and that it changes the overall direction only 
on a relatively large scale
(\href{https://youtu.be/oI3eCNtoPMA}
      {click for movie}).
}
\label{fig:U4_maze_exploring}
\end{figure}

We did let the robot evolve within the borders of a simple 
maze, as shown in Fig.~\ref{fig:U4_maze_exploring} and 
Suppl.~Video~4. Most of the time the robot is in the chaotic 
state, which is the dominant mode for the parameters used, 
namely $(w_0,z_0)=(180,80)$ and $U_{max}=4$. Intermittently,
after colliding with a wall, the robot switches to the
coexisting C1 mode. The radius of the stable C1 limit cycle 
in real-world coordinates is however so large, for 
$(w_0,z_0)=(180,80)$, that it does not fit into
the maze. The robot hence continues exploring. We
have obtained similar results when using a
$U_{max}=1$ chaotic mode.

A screenshot of a trajectory in open space is presented
in Fig.~\ref{fig:U4_chaos_screenshot}. One notices, that
the $U_{max}=4$ and $(w_0,z_0)=(180,80)$ chaotic mode 
wanders around aimlessly in much smother manner, than the 
$U_{max}=1$ chaotic mode shown in Fig.~\ref{fig:U1_chaos_screenshot}.
This is the result of topologically different attractor 
structures, as seen in the phase space of internal variables 
(see the supplementary materials).
Different types of chaos are indeed known to exist
\citep{wernecke2016partially}. 

The autonomous mode switching observed for the regular  
motion primitives can also be seen in Suppl.~Video~1.
For a detailed discussion of the possible switching 
scenarios see the supplementary materials.

\subsection{Switching between degenerate unstable limit cycles}

\begin{figure}[t]
\centering
\includegraphics[width=0.90\textwidth]{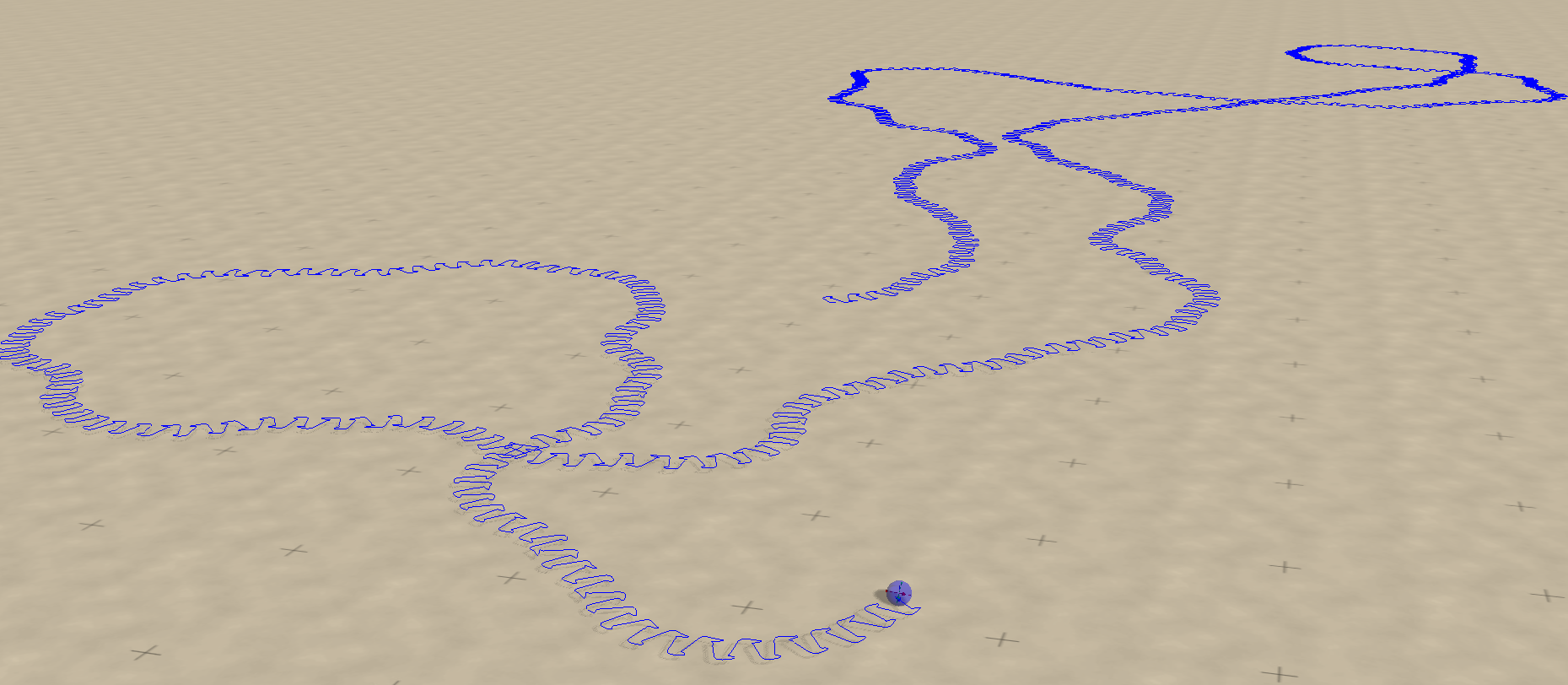}
\caption{Screenshot of the sphere robot in a chaotic 
mode for $U_{max}=4$ and $(w_0,z_0)=(180,80)$, indicated
by the arrow in Fig.~\ref{fig:U4_PD}. The blue line 
retraces the past trajectory. Note that the chaotic
wandering is substantially smoother than the one observed 
for the $U_{max}=1$ case (compare Fig.~\ref{fig:U1_chaos_screenshot}).
             }
\label{fig:U4_chaos_screenshot}
\end{figure}

\begin{figure}[t]
\centering
\includegraphics[width=0.9\textwidth]{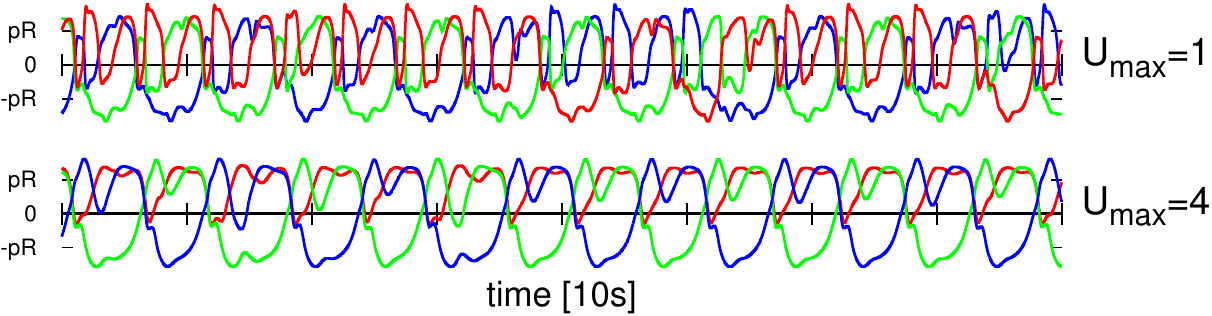}
\caption{As a function of time the positions of the three 
weights, compare Fig.~\ref{fig:three-axis-robot}, along 
the corresponding rods. 
{\it Top}: For the $U_{max}=1$ chaotic mode with 
$(w_0,z_0)=(210,400)$ shown in
Fig.~\ref{fig:U1_chaos_screenshot}.
{\it Bottom}: For the $U_{max}=4$ chaotic mode
$(w_0,z_0)=(180,80)$ shown in
Fig.~\ref{fig:U4_chaos_screenshot}.
Both modes are locally akin to an S2 mode,
albeit with substantial fluctuations (e.~g.~compare the 
bottom curvatures of the green line for $U_{max}=4$, see also 
Fig.~\ref{fig:modes_time_series}). Note that
phase slips do occur for the case of $U_{max}=1$, but
not for $U_{max}=4$.
}
\label{fig:modes_chaos_time_series}
\end{figure}

In Fig.~\ref{fig:modes_chaos_time_series} we compare
for the two chaotic modes, realized for $U_{max}=1$ 
and for $U_{max}=4$ respectively, the time series 
for the positions of the weights along the rods. One
observes, that the movements of the weight is
qualitatively similar, on short time scales, to 
an S2 mode (compare Fig.~\ref{fig:modes_time_series},
see also Suppl.~Video~3).
It is interesting, in this context, that the S2 mode 
has two types of degeneracies.
\begin{itemize}
\item Continuous. The S2 mode may propagate in any direction.
      There is hence a continuous manifold of attractors
      in the combined phase of controller, body and
      environment. Outside the actual region of stability
      this manifold contains either unstable limit
      cycles or limit cycle relicts \citep{gros2009cognitive}.
\item Discrete. There is a spontaneous symmetry breaking
      in the S2 mode, with two weights having identical
      but phase shifted movement patterns along their 
      respective rods, which are qualitatively different 
      to the trajectory of the third weight (see 
      Fig.~\ref{fig:modes_time_series}).
\end{itemize}
For the $U_{max}=4$ chaotic mode we did not observe
discrete mode switching, in above sense, which however
occurs frequently for the $U_{max}=1$ mode (see
Fig.~\ref{fig:modes_chaos_time_series}). The chaotic
meandering observed for the $U_{max}=4$ chaotic mode,
as evident in Fig.~\ref{fig:U4_chaos_screenshot}, is
hence a consequence of a smooth diffusion of the angle
of propagation on the manifold of unstable S2 limit
cycles (or limit cycle relicts \citep{linkerhand2013generating}).
In the phase space of the neural activity (as shown
in Fig.~5 of the supplementary material), the trajectory
corresponds to a chaotic phase diffusion along a limit
cycle \citep{wernecke2016partially}. This process is 
determinstic and not due to numerical errors, as we have 
checked by systematically reducting the stepsize used for 
the numerical integration. Noise is absent.

\section{Conclusions}

We have shown here, that a robot controlled by only 
a very limited number of neurons, three in our case,
may show complex behavior which may be interpreted
as explorative or playful. This is possible when
locomotion results from self-organizing processes
in the sensorimotor loop. The driving control dynamics, 
for which we have considered here short-term synaptic 
plasticity, then adapts itself seemingless to the physical 
requirements. No central controller is needed to detect an
external object \citep{rai2014learning},
or to switch direction when colliding 
with it. Stable and unstable limit cycles, together
with chaotic attractors, arise in the phase space of internal 
(control and robot body) variables. These attractors form
continua in the space of physical location and overall
propagation direction, with the chaotic locomotion
transitioning between unstable limit cycles. Transitions 
may either be between different types of regular locomotion, 
bounded circular or propagation meandering modes, or between 
the directions of unstable propagating limit cycles.

We note that the formation of a continuum of attractors is 
possible, whenever internal and external variables can be 
separated, such that internal variables span an independent 
subset of the phase space of the dynamical system.
Here, the position of the robot (on the ground plane, 
in the absence of obstacles) acts as an external variable, 
all the other variables being independent of it. 
The limit cycles and chaotic attractors, living in the 
subspace of internal variables, exist thus for all position 
vectors, generating a continuous degeneracy of locomotion modes.
The interactions with other robots and obstacles then results
in a transient breakdown of this degeneracy, which is restored 
instantaneously with the termination of physical contact.
Within this context, higher order control mechanisms would 
correspond to an external-variable dependent feedback, 
shaping the attractors either intermittently or slowly 
(with respect to the internal dynamics), thus leading possibly 
to the emergence of transiently stable attractors.

Our result, that the three-rod robot switches spontaneously
between a continuous set of attractors, in the chaotic state,
can be seen as a realization of chaotic wandering
\citep{tsuda2001toward}, which has been argued in turn
to occur in the brain in the form of self-organized
instabilities \citep{friston2012perception}, viz as
transient-state dynamics \citep{gros2007neural}.
There is furthermore a close relation to the concept
of attractor metadynamics \citep{gros2014attractor},
which denotes the either induced or spontaneous
switching between attracting sets.

The here simulated robot is furthermore compliant both on
the level of control and actuators, showing a highly flexible
response. The actuators are implemented by specifying
a target position for a limb, here a moving weight on
a rod. The force acting on the weight then results from
the interplay between the internal driving, provided
by a damped spring (between the actual and the target
position), with the physical restoring forces acting
on the weights, which in turn depend on the body
dynamics determined by the interaction with the ground,
obstacles and other robots \citep{floreano2014robotics}.

The isolated controlling network (realized in the limit
of infinitely strong actuators) can be interpreted in
addition as a central pattern generator \citep{steingrube2010self},
having a single intrinsic limit-cycle attractor.
The open-loop control incorporates however the feedback
of the environment through the induced forces. We find 
here, that the resulting embodiment \citep{cangelosi2015embodied} 
does morph the driving
dynamics of the central pattern generator not only quantitatively,
but also qualitatively, giving rise to a vast array of modes
which differ in part topologically from the dynamics of
the underlying central pattern generator. We believe that             
this dynamical systems approach of the locomotion of simple 
robots has not been fully exploited yet, having many 
interesting features and applications in store
for the field of neurorobotics.

\section*{Conflict of Interest Statement}

The authors declare that the research was 
conducted in the absence of any commercial 
or financial relationships that could be 
construed as a potential conflict of interest.

\section*{Author Contributions}
The experiments were conceived and designed by CG, BS and LM,
performed mainly by LM with BS adding some data.
The data was analyzed by CG, BS and LM, most of the 
plots produced by LM. The manuscript was mostly written 
by CG, with BS adding some paragraphs and revising it with LM.



\end{document}